\documentclass{article}

    \usepackage[nonatbib]{neurips_2021}

\usepackage[comma,authoryear]{natbib}
\newcommand{\RNum}[1]{\uppercase\expandafter{\romannumeral #1\relax}}
\usepackage[edges]{forest}
\usepackage[utf8]{inputenc} 
\usepackage[T1]{fontenc}    
\usepackage{hyperref}       
\usepackage{url}            
\usepackage{booktabs}       
\usepackage{amsfonts}       
\usepackage{nicefrac}       
\usepackage{microtype}      
\usepackage{amsmath,amsthm,amssymb,amsfonts}
\usepackage{amsmath,graphicx}
\usepackage{multicol}
 \usepackage{xcolor}
 \usepackage[export]{adjustbox}
\usepackage{xpatch}
\usepackage{hyperref}
\usepackage{comment}
   \usepackage{xcolor}
    \usepackage{lipsum}
\usepackage{booktabs,multirow,array}
    \usepackage{amsmath}
\usepackage{algorithm}
\usepackage{amsthm}
\makeatother

\usepackage{hyperref}

\AfterEndEnvironment{theorem}{\noindent\ignorespaces}

\AfterEndEnvironment{lemma}{\noindent\ignorespaces}

\usepackage{multirow}
\usepackage[noend]{algpseudocode}
\usepackage{amsthm}
\usepackage{enumitem}
\def\proj{\mathop{\rm proj}\nolimits}

\def\descent{\mathop{\rm descent}\nolimits}
\def\svdinit{\mathop{\rm svd-initialization}\nolimits}
\def\siteinit{\mathop{\rm site-initialization}\nolimits}
\usepackage{caption} 

\usepackage{caption,subcaption}
 \DeclareCaptionSubType[arabic]{figure}
 \usepackage{tikz}
\usetikzlibrary{arrows.meta,automata,positioning}
    \usepackage{mathtools}
\usepackage{amsmath,bm}
\usepackage[affil-it]{authblk}
\usepackage[english]{babel}

\usepackage{blindtext}
\usepackage{algorithm}
\usepackage{xcolor}
\usepackage[noend]{algpseudocode}
\def\proj{\mathop{\rm proj}\nolimits}

\def\trace{\mathop{\rm trace}\nolimits}
\def\descent{\mathop{\rm descent}\nolimits}

\usepackage{bbm}
\def\BState{\State\hskip-\ALG@thistlm}
\makeatother
\usepackage{bbm}
\usepackage{xpatch}

\makeatletter 
\usetikzlibrary{decorations.pathreplacing,calligraphy}

\NewDocumentCommand{\set}{o m}{%
  \IfNoValueTF{#1}
    {\{#2\}}
    {\{#1 \mid #2\}}%
}
\usepackage{tikz}
\usetikzlibrary{arrows.meta, 
                positioning,
                shadows}

\usepackage{amsmath,bm,times}
\title{Robust Hierarchical Patterns for identifying MDD patients: A Multisite Study}
\author[1,$\ast$]{Dushyant Sahoo\thanks{$^{\ast}$Corresponding author at: Center for Biomedical Image Computing and Analytic, Perelman School of Medicine,
   University of Pennsylvania, Philadelphia, USA \\
E-mail address: sadu@seas.upenn.edu}}
\author[1]{Mathilde Antoniades}
\author[2,3]{Cynthia H.Y. Fu}
\author[1]{Christos Davatzikos}
\affil[1]{Center for Biomedical Image Computing and Analytic, Perelman School of Medicine,
   University of Pennsylvania, Philadelphia, USA}
\affil[2]{School of Psychology,
    University of East London, 
    London, UK}
\affil[3]{     Centre for Affective Disorders, Institute of Psychiatry, Psychology and Neuroscience, King's College London, London, UK}
\date{}                     

\begin{document}
\nolinenumbers
\maketitle

\begin{abstract}
Many supervised machine learning frameworks have been proposed for disease classification using functional magnetic resonance imaging (fMRI) data, producing important biomarkers. More recently, data pooling has flourished, making the result generalizable across a large population. But, this success depends on the population diversity and variability introduced due to the pooling of the data that is not a primary research interest. Here, we look at hierarchical Sparse Connectivity Patterns (hSCPs) as biomarkers for major depressive disorder (MDD). We propose a novel model based on hSCPs to predict MDD patients from functional connectivity matrices extracted from resting-state fMRI data. Our model consists of three coupled terms. The ﬁrst term decomposes connectivity matrices into hierarchical low-rank sparse components corresponding to synchronous patterns across the human brain. These components are then combined via patient-speciﬁc weights capturing heterogeneity in the data. The second term is a classification loss that uses the patient-specific weights to classify MDD patients from healthy ones. Both of these terms are combined with the third term, a robustness loss function to improve the reproducibility of hSCPs. This reduces the variability introduced due to site and population diversity (age and sex) on the predictive accuracy and pattern stability in a large dataset pooled from five different sites. Our results show the impact of diversity on prediction performance. Our model can reduce diversity and improve the predictive and generalizing capability of the components. Finally, our results show that our proposed model can robustly identify clinically relevant patterns characteristic of MDD with high reproducibility.
\end{abstract}

\section{Introduction}
MDD is one of the most widespread psychiatric disorders characterized by persistent sadness, depressed mood, low self-esteem, sleep disturbances, emotional changes, and loss of interest in pleasurable activities, causing disruptions to daily life \cite{belmaker2008major}. In addition, MDD causes more than $800,000$ deaths each year globally and is also the leading cause of disability \cite{otte2016major}. Understanding the mechanism of MDD is crucial for effective diagnosis, treatment and prevention, and understanding the functioning of the human brain in a depressive state compared to a healthy one. Considering the breadth of symptoms of this disease, it follows that disruptions within and across multiple brains systems and networks must be at play. Indeed, much work has been done to understand the functional brain changes associated with MDD. However, much remains unknown about the pathophysiology of the disease and the rates of relapse and recurrence remain high \cite{mueller1999recurrence,kessler2012costs}.

Previous studies have shown that MDD is associated with disruptions in regional functional connectivity and abnormal functional integration of distributed brain regions \cite{greicius2007resting,liu2013abnormal,wu2011abnormal,zhu2012evidence}. More recent approaches using seed-based connectivity, independent component analyses, network homogeneity and graph theory for functional connectivity analyses have revealed similar findings- disruptions in functional networks and in between functional networks across specific region pairs in MDD. The brain networks exhibiting abnormal interactions in MDD include the Default Mode Network (awareness of internal states), Dorsal Attention Network (external awareness), Fronto-Parietal Network (top-down regulation of attention and emotion), Salient Network (salient events) and Affective Network (emotion processing) \cite{ye2015changes,kaiser2015large,yan2019reduced,mulders2015resting,iwabuchi2015localized,brakowski2017resting}. 

Many efforts have been made to build functional connectivity-based predictive models for identifying network-based biomarkers of depression \cite{craddock2009disease,zeng2012identifying,bhaumik2017multivariate,rosa2015sparse,zhao2020functional}. Majority of studies are based on multivariate pattern analysis of functional connectivity. Dictionary learning is one such approach that can model low rank, sparse group-level interpretable networks having high predictive power \cite{batmanghelich2011generative,eavani2013unsupervised}.  \citet{eavani2015identifying} proposed sparse Connectivity Patterns (SCPs), which decompose the connectivity matrix of each subject as a set of shared sparse interpretable patterns and subject-specific information. They extended the method by adding a discriminative term to classify young adults vs. children \cite{eavani2014discriminative}. More recently, \citet{d2020joint} used SCPs based generative-discriminative model to predict clinically relevant networks characteristic of Autism Spectrum Disorder.

Accurately classifying case-control groups has proven challenging due to inter-patient variability \cite{benkarim2021cost}. In addition, biomarkers and brain patterns learned for prediction from small and homogeneous datasets are difficult to replicate and poorly generalize to new cohorts. In order to improve generalizability and replicability of brain patterns and evaluate a hypothesis in multiple sites/settings, there have been several open-access neuroimaging data-sharing initiatives \cite{alexander2017open,biswal2010toward,casey2018adolescent,di2017enhancing}. In these initiatives, data is pooled from multiple sites to capture demographically diverse populations therefore building heterogeneous datasets that are more likely to reflect the wider population. 

Surprisingly, multi-site studies have shown lower classification performance and poor generalization to data from new cohorts compared to single-site studies \cite{arbabshirani2017single,chen2016multivariate,nielsen2013multisite}. Analyzing data from these initiatives poses an inherent challenge due to variability introduced from the diverging backgrounds of the subjects and from site differences in MRI scanner hardware and software \cite{kostro2014correction,yamashita2019harmonization,shinohara2017volumetric,abraham2017deriving,jovicich2016longitudinal,noble2017multisite}. The non-biological variability introduced due to pooling of the data can affect the biomarkers or common features extracted from fMRI data \cite{yu2018statistical}, these include functional connectivity \cite{shinohara2017volumetric} and sparse hierarchical factors \cite{sahoo2021learning}. The non-biological variability can lead to decreased statistical power, spurious results and difficulty in identifying robust biomarkers depending on the task. In addition, the correlation between site effects and biological predictors can lead to an incorrect inference of non-biological differences as biological. Thus, many neuroimaging studies need to develop robust models that remove the non-biological variance and extract biologically relevant information.

We use hierarchical sparse connectivity patterns (hSCPs) \cite{sahoo2020hierarchical} to extract the relation between various functional networks in the human brain. The method proposes a correlation matrix decomposition strategy, capturing sparse hierarchical components agnostic to the parcellation scheme. The sparse representation helps identify shared co-activation patterns in all the subjects while subject-specific coefficients model the dataset's heterogeneity. See Appendix \ref{sec:rel_work} for extended discussion on related work. The hSCP can find relations between various functional networks across various populations, which can help in discovering functional incoherence caused due to MDD.

This paper aimed to identify the hSCP biomarkers of MDD while reducing the effects of diversity (age, sex, site) common to large pooled datasets. Our work builds on the hSCP model by using the discriminative nature of the subject-specific weights extracted from hSCPs to classify MDD via logistic regression in a large multi-site study. To tackle the variability introduced due to pooling of the datasets, we use robust to site hSCP (rshSCP) \cite{sahoo2021learning}. The rshSCP method captures linear site effects and uses adversarial learning to reduce site effects in the subject-specific coefficients. We extend robust to site hSCP and introduce discriminative rshSCP (dis-rshSCP) to extract homogeneous components discriminative of MDD by reducing heterogeneity introduced due to covariates (age, sex and site), which are known to affect neuroimaging analysis \cite{alfaro2021confound,duncan2013overview}. Experiments on real datasets show that our approach can improve the split-sample and leave one site predictability power of the components while retaining the reproducibility of the components, thus capturing informative heterogeneity. The classification performance on unseen data indicates the generalizability of the model. Our results demonstrate that MDD is associated with increased and decreased representation in patterns associated with various functional networks. The results demonstrate our framework's potential in identifying patient-predictive biomarkers of a MDD. 

\textbf{Outline:} We start by reviewing hierarchial Sparse Connectivity Patterns (hSCPs). Then, in Section  \ref{sec:method_robust}, we present our method, extracting interpretable hSCPs which are discriminative of MDD and are robust to covariates (age, sex and site). In Section \ref{sec:robust_real_dataset}, we demonstrate that our method could extract hSCPs with high reproducibility and prediction power. This is followed by a discussion on the interpretability of the extracted patterns, limitations and future work.


\section{Preliminaries}
We first introduce preliminaries that serve as a basis for our discussions.

\paragraph{Notations:} We follow the same notation as in \citet{sahoo2020hierarchical,sahoo2021extraction,sahoo2021learning}. $\mathbb{S}^{P \times P}_{++}$ denotes symmetric positive definite matrices of size $P \times P$. $\mathbf{A} \geq 0$ denotes that all the elements of matrix $\mathbf{A}$ are greater than or equal to $0$. $P \times P$ matrix with all elements equal to one is denoted by $\mathbf{J}_P$. Identity matrix of size $P \times P$ is denoted by $\mathbf{I}_P$ and $\mathbf{A} \circ \mathbf{B}$ denotes element-wise product between two matrices $\mathbf{A}$ and $\mathbf{B}$.

\paragraph{Problem Setup:} The fMRI data of the $i^{th}$ subject having $P$ regions and $T$ time points is denoted by $\mathbf{X}^i \in \mathbb{R}^{P \times T}$ with total $N$ number of subjects or participants. Let $\mathbf{\Theta}^i \in \mathbb{S}^{P \times P}_{++}$ be the correlation matrix where $\mathbf{\Theta}^i_{m,o}$ stores the correlation between time series of $m^{th}$ and $o^{th}$ node. Let there be total $S$ sites in the multi-site data and $\mathcal{I}_s$ be the set storing subjects from site $s$. Let $\mathbf{y}_{site} \in \mathbb{R}^{N \times S}$ be site labels encoded in one-hot manner, and $\mathbf{y}_{age} \in \mathbb{R}^{N}$, $\mathbf{y}_{sex} \in \mathbb{R}^{N}$ and $\mathbf{y}_{mdd} \in \mathbb{R}^{N}$ be the vectors storing information about age, sex and MDD label. We aim to extract set of hierarchical patterns representative of depression using fMRI data with reduced variability introduced due to age, sex and site. 

\begin{figure}
\centering

\begin{tikzpicture}

\begin{scope}
\coordinate (A) at (0,0);
\coordinate (B) at (0,6);
\coordinate (C) at (12.2,6);
\coordinate (D) at (12.2,3);
\coordinate (E) at (4,3);
\coordinate (F) at (4,0);
 \path[draw,clip,decorate,rounded corners]  (A) to (B) to (C)
         to (D)
         to (E)
         to (F)
         to cycle;

\fill[blue!20] (A) circle (180mm);
\node[inner sep=0pt] (russell) at (2,4)
    {\includegraphics[width=.15\textwidth]{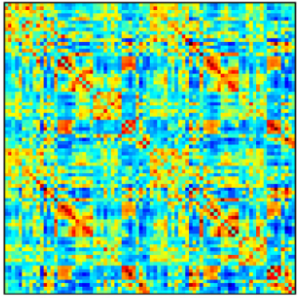}};
    
\node[inner sep=0pt] (whitehead) at (5,4)
    {\includegraphics[width=.007\textwidth]{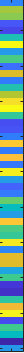}};
    \node[inner sep=0pt] (whitehead) at (5.2,4)
    {\includegraphics[width=.0065\textwidth]{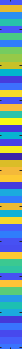}};
    
\node[inner sep=0pt] (whitehead) at (11,4)
    {\includegraphics[width=.007\textwidth]{Figures/Images/c1.png}};
    \node[inner sep=0pt] (whitehead) at (11.2,4)
    {\includegraphics[width=.0065\textwidth]{Figures/Images/c2.png}};    
   
 \node[inner sep=0pt] (whitehead) at (2,1)
    {\includegraphics[width=.28\textwidth]{ 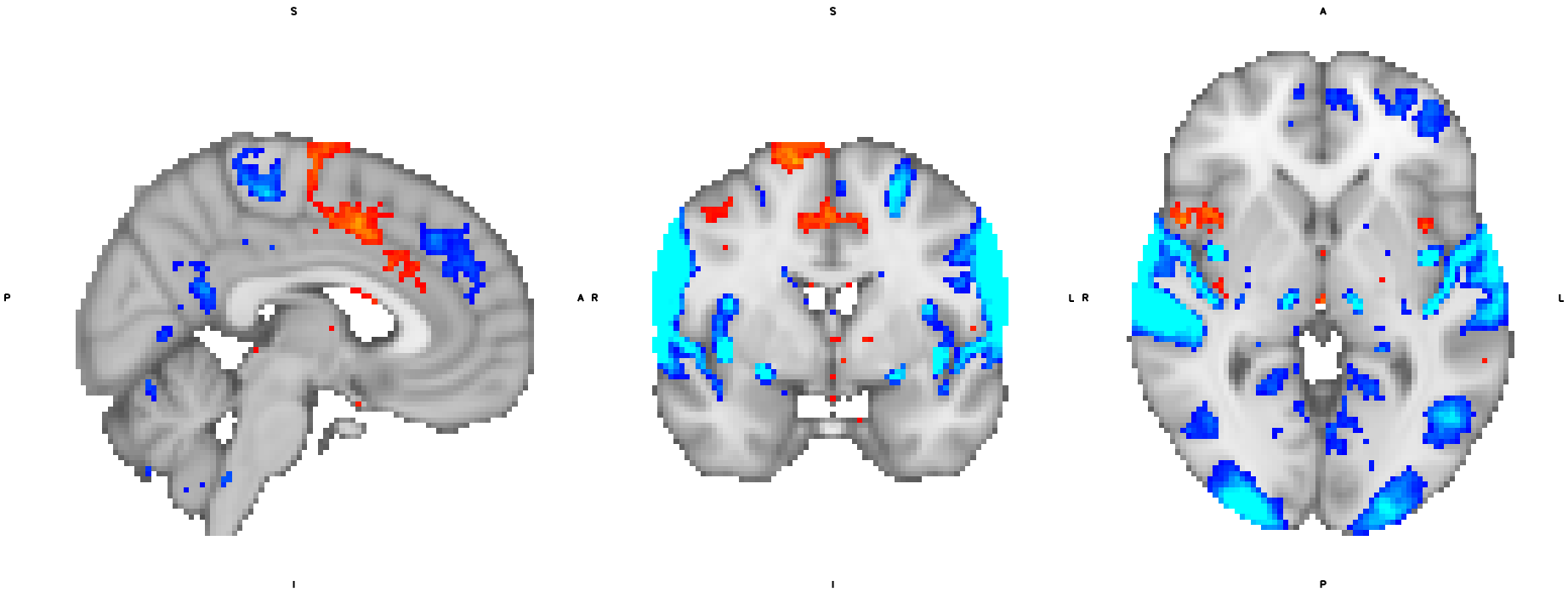}};  
     \node[inner sep=0pt] (whitehead) at (8.2,4)
    {\includegraphics[width=.15\textwidth]{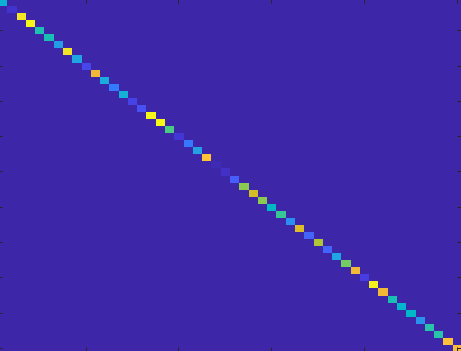}};

    \draw [decorate,
    decoration = {brace}] (4.9,3.2) --  (4.9,4.8);
    \draw [decorate,
    decoration = {brace,mirror}] (5.7,3.2) --  (5.7,4.8);
     \node[text width=3cm] at (6.8,4) 
    {$\cdots$};
    
    \draw [decorate,
    decoration = {brace}] (10.9,3.2) --  (10.9,4.8);
    \draw [decorate,
    decoration = {brace,mirror}] (11.7,3.2) --  (11.7,4.8);
     \node[text width=3cm] at (12.8,4) 
    {$\cdots$};
    
     \node[text width=3cm] at (3.4,5.5) 
    {$\mathbf{\Theta}^n$};
    \node[text width=3cm] at (5.4,4) 
    {$=$};
     \node[text width=3cm] at (6.6,5.5) 
    {$\mathbf{W}$};
    \node[text width=3cm] at (7.8,4) 
    {$*$};
     \node[text width=3cm] at (9.6,5.5) 
    {$\mathbf{\Lambda}^n$};
        \node[text width=3cm] at (11.4,4) 
    {$*$};
         \node[text width=3cm] at (12.6,5.5) 
    {$\mathbf{W}^\top$};    
    \node[text width=3cm] at (13.3,4.8) 
    {$\top$};    
    
\end{scope}

\begin{scope}
\coordinate (A1) at (4.3,0);
\coordinate (B1) at (4.3,2.7);
\coordinate (C1) at (12.2,2.7);
\coordinate (D1) at (12.2,0);

 \path[draw,clip,decorate,rounded corners]  (A1) to (B1) to (C1)
         to (D1)
         to cycle;
\fill[green!20] (A1) circle (140mm);

\node[inner sep=0pt] (whitehead) at (5.5,1.3)
 {\includegraphics[width=.15\textwidth]{ 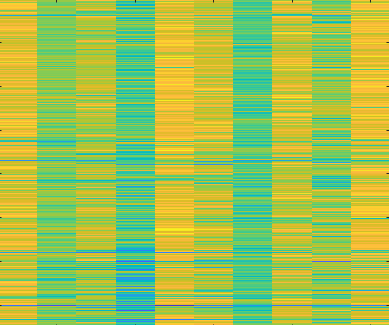}}; 
 
 \node[inner sep=0pt] (whitehead) at (8.3,1.3)
 {\includegraphics[width=.01\textwidth]{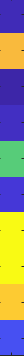}}; 
 
 \node[inner sep=0pt] (whitehead) at (11.3,1.3)
 {\includegraphics[width=.008\textwidth]{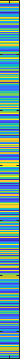}}; 

     \node[text width=3cm] at (6.7,2.5) 
    {$\mathcal{D}$};
     \node[text width=3cm] at (9.7,2.5) 
    {$\mathbf{b}$};    
         \node[text width=3cm] at (12.7,2.5) 
    {$\mathbf{y}_{mdd}$};  
\node[text width=3cm] at (8.6,1.3) 
    {$*$};
\node[text width=3cm] at (10.8,1.3) 
    {$\xrightarrow{f(\mathcal{D},\mathbf{b})}$};

\end{scope}

\begin{scope}[very thick,->]
  \draw (4.4,3.4)--(3.4,2.4);

  \draw (6.8,3.4)--(5.8,2.4);
\end{scope}

\begin{scope}
  \node[text width=5cm,rotate=90] at (-0.5,3.3) 
    {\textcolor{blue!40}{Robustly Learning Components}}; 
     \node[text width=4cm,rotate=270] at (12.5,1) 
    {\textcolor{green}{Classification Model}}; 
\end{scope}

\end{tikzpicture}

\caption{A joint two level modeling for connectivity analysis and prediction. First part is functional data representation depicted in blue box. Here the correlation matrices are decomposed into shared components stored in a basis matrix and subject specific information. We learn this decomposition in a robust manner to reduce variability due to site and demographics. To visualize each component, a column of basis matrix is projected onto the brain. Second part is the prediction of MDD patients depicted in green box. } 
\label{fig:ill}
\end{figure}

\subsection{Introduction to hierarchical Sparse Connectivity Patterns}
\label{sec:hSCP}
A graphical summary of our model is presented in Fig. \ref{fig:ill}. The two inputs to our model are the rs-fMRI correlation matrices (upper left) and the binary scores depicting if a person is healthy or has MDD (lower right). The correlation matrices are constructed from the time series data describing the similarity using Pearson's Correlation Coefﬁcient between various nodes of the human brain. The blue box in Fig \ref{fig:ill} indicates the generative model estimating the components. Here, we decompose correlation matrices into a set of components capturing co-activation patterns common across the entire cohort and subject-specific information capturing heterogeneity in the data representing the strength of each component in each individual. We capture this information while reducing the effects of demographics and site to improve the generalizability of the components and predictability of the subject-specific information. The green box indicates the discriminative model guiding the components to represent MDD. Here, we use the information from the subject-speciﬁc coefﬁcients to predict MDD via a classification model for each individual. $f()$ takes subject-specific information and a set of weights as input and maps it to a binary value representing $1$ and $0$ for MDD and healthy subjects.

Hierarchical Sparse Connectivity Patterns (hSCP) \cite{sahoo2020hierarchical} is an extension of Sparse Connectivity Patterns (SCPs) \cite{eavani2015identifying} to find hierarchical structures in the fMRI data to have multi-scale representation of connectivity patterns in the human brain. It was first defined by \cite{eavani2015identifying} to estimate sparse co-activated patterns in the human brain. As seen in Fig. \ref{fig:ill}, we model $\mathbf{\Theta}^n$ using a shared representation and a subject-speciﬁc term storing information about each component's strength. The $\mathbf{W}_1 \in \mathbb{R}^{P \times k_1}$ stores $k_1$ is a concatenation of $K$ components $\mathbf{w}_l \in \mathbb{R}^{P \times 1}$ , i.e.
$\mathbf{W} \coloneqq \left[\mathbf{w}_1 \; \mathbf{w}_2 \cdots \mathbf{w}_k \right]$, where $k \leq P$. These components capture steady-state co-activation patterns across regions in the brain shared across all the population. While components are shared across all the subjects, their representation in each subject is unique, captured by $\mathbf{\Lambda}_1^n$. $\mathbf{\Lambda}_1^n$ is kept to be non-negative to ensure the positive semi-definite structure of the correlation matrices $\mathbf{\Theta}^n$. The SCP representation is written as:
\begin{align*}
    \mathbf{\Theta}^n \approx \mathbf{W}_1\mathbf{\Lambda}_1^n \mathbf{W}_1^\top.
\end{align*}

This formulation helps reduce the high dimensionality of the data by projecting it into lower-dimensional space and providing subject-level information. SCP can be extended to hSCP following the below equations:
\begin{equation}
 \begin{aligned}
\mathbf{\Theta}^n &\approx \mathbf{W}_1\mathbf{\Lambda}_1^n \mathbf{W}_1^\top, \\ \mathbf{\Theta}^n &\approx \mathbf{W}_1\mathbf{W}_2\mathbf{\Lambda}_1^n \mathbf{W}_2^\top \mathbf{W}_1^\top, \\
& \vdots \\\mathbf{\Theta}^n &\approx \mathbf{W}_1\mathbf{W}_2 \ldots\mathbf{W}_K \mathbf{\Lambda}^n_K \mathbf{W}_K^\top\mathbf{W}_{K-1}^\top \ldots \mathbf{W}_1^\top ,
 \end{aligned}
 \label{eq:hscp}
 \end{equation}
 where $K$ is the level of hierarchy in the formulation, $\mathbf{\Lambda}^n_k$ is a non-negative diagonal matrix storing relative contribution of each component for $n$th subject at $k$th level and $P > k_1 > \ldots > k_K $ to make the formulation low dimensional and forcing each successive level in the hierarchy to have less number of components than the previous level. In the hierarchical structure, at the bottom level we have $\mathbf{W}_1 \in \mathbb{R}^{P \times k_1}$ storing $k_1$ components. These components are then linearly transformed at each successive level to get lower-dimensional components than the previous level. $\mathbf{W}_2$, $\mathbf{W}_3$, $\ldots$, $\mathbf{W}_K$ are used to transform $k_1$ dimensional components to $k_2, k_3, \ldots, k_K$ dimensional components where $P > k_1 > \ldots > k_K $. For example, let $\mathbf{W}_1$ be the components at lower most level, then to extract components at level above this, we multiply $\mathbf{W}_1$ by $\mathbf{W}_2$ to obtain $k_2$ dimensional components which are extracted using $k_1$ dimensional components. All the elements of $\mathbf{W}_2$, $\mathbf{W}_3$, $\ldots$, $\mathbf{W}_K$ are constrained to be greater than $0$ for interpretability. We now define sets storing the correlation matrices, components and weights at all level for notational brevity. Set of correlation matrices is denoted by $\mathcal{C} = {\set[\mathbf{\Theta}^n]{n=1,\ldots, N}}$, set storing information about shared components is denoted by $\mathcal{W} = {\set[\mathbf{W}_r]{r=1,\ldots, K}}$ and the set storing subject-specific information is denoted by $\mathcal{D}= {\set[\mathbf{\Lambda}_r^n]{r=1,\ldots, K; n=1\ldots,N}}$ with $\mathbf{\Lambda}^n_r \geq 0$. The hierarchical components and subject-specific information are estimated by performing alternating minimization on the below optimization problem:
  \begin{equation}
\begin{aligned}
& \underset{\mathcal{W},\mathcal{D}}{\text{min}}
& & \sum_{n=1}^{N} \sum_{r=1}^{K}\|\mathbf{\Theta}^n - (\prod_{j=1}^{r}\mathbf{W}_j)\mathbf{\Lambda}_r^n(\prod_{j=1}^{r}\mathbf{W}_n)^\top \|_F^2\\
& \text{s.t.}
&&\|\mathbf{w}^r_l\|_1 < \lambda_r, \; l=1,\ldots,k_r; \; r=1,\ldots,K\\ 
&&& \|\mathbf{w}^r_l\|_\infty \leq 1, \; l=1,\ldots,k_r; \; r=1,\ldots,K\\
&&& \trace(\mathbf{\Lambda}_r^n) =1,  \; n=1,\ldots,N; \; r=1,\ldots,K\\
&&&\mathbf{\Lambda}_r^n \geq 0, \; \; n=1,\ldots,N; \; r=1,\ldots,K \\
&&&\mathbf{W}_j \geq 0, \; j=2,\ldots,K.
\end{aligned}
   \label{problem:hscp}
 \end{equation}
The first constraint in the above optimization problem is $L_1$ which helps in capturing compact and clinically informative representation by encouraging sparsity in the components. $L_1$ regularizer will help in selecting a small number of nonzero entries in $\mathbf{W}$ that explain the data. Identifiability and reproducibility issues are solved by combining $L_1$, $L_{\infty}$ and $\trace$ constraints. $\mathbf{\Lambda}_r^n \geq 0$ constraint ensures that the subject-specific information is kept positive, and therefore the reconstructed matrix is positive semi-definite. Again for notational brevity, we will use the set notation for the above constraints by denoting $ \Omega_{\mathcal{W}} = \set[\mathbf{W}]{\|\mathbf{w}^r_l\|_1 < \lambda_r,\; \|\mathbf{w}^r_l\|_\infty \leq 1,  \mathbf{W}_j \geq 0, \; j=2,\ldots,K}  $ and $\Psi = \set[\mathbf{\Lambda}]{\trace(\mathbf{\Lambda}_r^n) =1,\mathbf{\Lambda}_r^n \geq 0}$. Refer to \citet{sahoo2020hierarchical} for a detailed description of the method.

\subsection{Reducing site effects}
In this section, we discuss framework proposed by \citet{sahoo2021extraction} to reduce site effects in the hSCP model. Their idea is to jointly model sparse patterns and site information, and use adversarial learning to reduce site effects. We first look at the case with one level in the hierarchy for easier understanding. This will extended to multiple levels in the upcoming section. Authors of the paper hypothesize that site and scanner information for the training data can be represented by $\mathbf{V} \in \mathbb{R}^{P \times P}$ space. For each site $s=1,\ldots, S$, let there be diagonal matrices $\mathbf{U}^s \in \mathbb{R}^{P \times P}$ storing additive and $\mathbf{Z}^s \in \mathbb{R}^{k \times k}$ storing scaled site-specific information. Based on the above hypothesis, $\mathbf{\Theta}^{n}$ is decomposed to jointly estimate the hSCPs, site and scanner information for $n \in  \mathcal{I}_s$ as:
\begin{align}
    \mathbf{\Theta}^{n} \approx\mathbf{W} \mathbf{Z}^s \mathbf{\Lambda}^{n} \mathbf{W}^\top + \mathbf{U}^s\mathbf{V}. 
    \label{eq:site}
\end{align}
The method imposes $ L_1$ sparsity constraint on $\mathbf{V}$ to prevent overfitting.

 \section{Method}
 \label{sec:method_robust}
 \subsection{Robust to covariates hSCP}
In the previous section, we saw how site and scanner effects could be stored using $\mathbf{U}$ and $\mathbf{V}$. \citet{sahoo2021extraction} added an adversarial layer on top of it to reduce the predictive power of $\mathcal{D}$ storing subject-specific coefficients for predicting site. In addition to reducing site effects in this paper, we also minimize variability due to age and sex. For this, we train a model $F(\zeta, \mathcal{D})$ parameterized by $\zeta$ with input $\mathcal{D}$ that return age $\hat{\mathbf{y}}_{age} \in \mathbb{R}^{N}$, sex $\hat{\mathbf{y}}_{sex} \in \mathbb{R}^{N}$ and site $\hat{\mathbf{y}}_{site} \in \mathbb{R}^{N \times S}$ predictions. We train the model by minimizing the below loss function $\mathcal{L}_{1}(\zeta, \mathcal{D}, \mathbf{y}_{site},\mathbf{y}_{age},\mathbf{y}_{sex})$:
\begin{equation}
\begin{aligned}
    \mathcal{L}_{1} = \underbrace{\|\hat{\mathbf{y}}_{age} - {\mathbf{y}}_{age} \|_2^2}_{\substack{\text{preserve age} \\ \text{information}}} -\sum_{s=1}^S \sum_{n=1}^N  \underbrace{y^{n,s}_{site}\log \hat{y}^{n,s}_{site}}_{\substack{\text{preserve site} \\ \text{information}}}   -\sum_{n=1}^N \underbrace{\left(y^{n}_{sex}\log \hat{y}^{n}_{sex}+(1-y^{n}_{sex})\log (1-\hat{y}^{n}_{sex}) \right)}_{\substack{\text{preserve sex} \\ \text{information}}}.
    \end{aligned}
\end{equation}
Let $\zeta^* =  \underset{\zeta}{\text{arg min}} \; \mathcal{L}_{1}$ be optimum value which minimizes $\mathcal{L}_{1}$. The above problem is the Multi-task learning (MTL) \cite{caruana1997multitask,zhang2017survey} problem to learn multiple correlated tasks at the same time. This formulation helps improve the performance of each and reduces the need to introduce multiple models to solve individual tasks. We use the direct sum approach to combine different objectives by directly minimizing the sum of all losses of different tasks, a common practice in multi-task learning. A hard parameter sharing strategy is used to model network architecture. In this strategy, parameters are shared by the bottom layers among all tasks, while top layers are selected to be task-specific, helping with robustness against overfitting \cite{ruder2017overview}. It is a commonly used method for designing deep learning models in the literature \cite{long2015learning,ruder2019latent,sener2018multi}. Details about the architecture are given in Appendix \ref{sec:mtl}.

Using the loss function $\mathcal{L}_{1}$, we modify $\mathbf{\Lambda}^{n}$ such that its predictability power to predict site, age and sex reduces. This can be achieved by maximizing $\mathcal{L}_{1}$ loss with respect to $\mathbf{\Lambda}^{n}$. Note that we are trying to solve two problems with one loss function, first is finding optimal $\zeta$ which minimizes $\mathcal{L}_{1}$ and second is finding optimal $\mathbf{\Lambda}^{n}$ which maximizes $\mathcal{L}_{1}$. This will result in a minimax game, where the $\zeta$ is learned to minimize the cross-entropy and regression loss, and $\mathbf{\Lambda}^{n}$ is adjusted to maximize the loss. The minimax optimization problem can be written as:
 \begin{equation}
\begin{aligned}
\underset{\zeta}{\text{max}} \; &\underset{\mathbf{W},\mathcal{D},\mathcal{U},\mathbf{V},\mathcal{Z}}{\text{min}} \quad && \sum_{s=1}^S \sum_{n \in \mathcal{I}_s}   \|\mathbf{\Theta}^{n} - \mathbf{W} \mathbf{Z}^s\mathbf{\Lambda}^{n} \mathbf{W}^\top - \mathbf{U}^s\mathbf{V}\|_F^2 - \gamma_1 \mathcal{L}_{1}(\zeta, \mathcal{D}, \mathbf{y}_{site},\mathbf{y}_{age},\mathbf{y}_{sex}) \\
& \hspace{1em} s.t. && \mathbf{W} \in \Omega, \quad \mathcal{D} \in \Psi, \quad \|\mathbf{v}_p\|_1 < \mu, \; p=1,\ldots,P,
\end{aligned}
\label{eq:site_ind}
\end{equation}
where $\mathcal{U} = \{\mathbf{U}_s|s=1,\ldots,S \}$, $\mathcal{Z} = \{\mathbf{Z}_s|s=1,\ldots,S \}$ and $\mathbf{v}_p$ is the $p$th column of $\mathbf{V}$. Here $\|\mathbf{\Theta}^{n} - \mathbf{W} \mathbf{Z}^s\mathbf{\Lambda}^{n} \mathbf{W}^\top - \mathbf{U}^s\mathbf{V}\|_F^2$ is the total error in the representation of the $S$ subjects and $\mathcal{L}_{1}$ is the robustness loss, and $\gamma_1$ is the tradeoff between representation learning and robustness.

 \subsection{Joint modeling of MDD scores}
 The aim of the paper is to learn components which are representative of MDD. For this, we build discriminative rshSCP (dis-rshSCP) to use subject-specific information $\mathbf{\Lambda}^n_r$ to predict whether subject $n$ has MDD or not. Use this model we will subject-specific information which is most predictive of MDD and will give components corresponding to that. We model MDD information using logistic regression framework with parameter $\mathbf{b} \in \mathbb{R}^{f}$, where $f = \sum_{r=1}^K k_r$, subject specific information of all the levels combined $\mathbf{t}^n = diag[\mathbf{\Lambda_1^n},\ldots,\mathbf{\Lambda_K^n}]$  and the loss function defined below:
  \begin{equation}
\begin{aligned}
  \mathcal{L}_{2}(\mathbf{b}, \mathcal{D}, \mathbf{y}_{mdd}) = - \sum_{n=1}^N \left[y^n_{mdd} \log\left(\frac{1}{1+\exp{(-\mathbf{b}^\top \mathbf{t}^n)}} \right) + (1- y^n_{mdd}) \log\left(\frac{\exp{(-\mathbf{b}^\top \mathbf{t}^n)}}{1+\exp{(-\mathbf{b}^\top \mathbf{t}^n)}} \right) \right].
\end{aligned}
\label{eq:logistic}
\end{equation}
 We minimize cross entropy loss $\mathcal{L}_{2}(\mathbf{b}, \mathcal{D}, \mathbf{y}_{mdd})$ with $\mathbf{b}$ as the parameters to be estimated. Now, let the generative loss be 
\begin{align}
G(\mathcal{C},\mathcal{W},\mathcal{D},\mathcal{U},\mathbf{V},\mathcal{Z}) =   \sum_{s=1}^S \sum_{n \in \mathcal{I}_s} \sum_{r=1}^{K}\|\mathbf{\Theta}^n - (\prod_{j=1}^{r}\mathbf{W}_j)\mathbf{Z}^s_r \mathbf{\Lambda}_r^n(\prod_{j=1}^{r}\mathbf{W}_n)^\top - \mathbf{U}^s_r \mathbf{V}_r \|_F^2,     
\end{align}
which models the components and site information, then the joint optimization problem can be written as: 
 \begin{equation}
\begin{aligned}
\underset{\zeta}{\text{max}} \; &\underset{ \mathcal{W},\mathcal{D},\mathcal{U},\mathcal{Z},\mathbf{V},\mathbf{b}}{\text{min}} \quad &&  \underbrace{G(\mathcal{C},\mathcal{W},\mathcal{D},\mathcal{U},\mathbf{V},\mathcal{Z})}_{\substack{\text{learn subject and} \\ \text{site information}}} - \gamma_1 \underbrace{\mathcal{L}_{1}(\zeta, \mathcal{D}, \mathbf{y}_{site},\mathbf{y}_{age},\mathbf{y}_{sex})}_{\substack{\text{reduce age, sex and} \\ \text{site information}}} + \gamma_{2} \underbrace{\mathcal{L}_{2}(\mathbf{b}, \mathcal{D}, \mathbf{y}_{mdd})}_{\substack{\text{preserve MDD} \\ \text{information}}} \\
& \hspace{1em} s.t. && \mathcal{W} \in \Omega, \quad \mathcal{D} \in \Psi, \quad \|\mathbf{v}_p\|_1 < \mu, \; p=1,\ldots,P.
\end{aligned}
\label{eq:joint_mdd}
\end{equation}
Here, in addition to representation learning and robustness loss, we also have prediction error $\mathcal{L}_2$. $\gamma_1$ and $\gamma_2$ are the trade-offs between representation learning, robustness, and prediction.

\subsection{Prediction on unseen data}
To estimate $\mathbf{\Lambda}$ for a new subject, we first solve the optimization problem in equation \ref{eq:site_ind} to estimate $\mathbf{W}$ computed from the training data. The estimation of the coefficients of unseen subjects are then estimated by solving the below minimax problem where $\mathbf{W}$ is computed from the training data:
 \begin{equation}
\begin{aligned}
\underset{\zeta}{\text{max}} \; &\underset{\mathcal{D},\mathcal{U},\mathcal{Z},\mathbf{V}}{\text{min}} \quad &&  G(\mathcal{C},\mathcal{W},\mathcal{D},\mathcal{U},\mathbf{V},\mathcal{Z}) - \gamma_1 \mathcal{L}_{1}(\zeta, \mathcal{D}, \mathbf{y}_{site},\mathbf{y}_{age},\mathbf{y}_{sex}) \\
& \hspace{1em} s.t. &&  \mathcal{D} \in \Psi, \quad \|\mathbf{v}_p\|_1 < \mu, \; p=1,\ldots,P.
\end{aligned}
\label{eq:test}
\end{equation}
The estimate for the MDD information for the test subject $n$ is given by:
   $$  y_{mdd}^n = \begin{cases*}
                    1 & \text{if  $\frac{1}{1+ \exp{(-\mathbf{b}^\top \mathbf{t}^n)}} \ge 0.5$ } \\
                     0 & \text{otherwise},
                 \end{cases*} $$%
                
where $\mathbf{b}$ and $\mathbf{t}$ is estimated from solving equation \ref{eq:joint_mdd} and \ref{eq:test} respectively. 

The optimization problems defined in \ref{eq:joint_mdd} and above \ref{eq:test} are non-convex problems. We use alternating minimization to solve, Appendix \ref{sec:alg} gives the detail of the complete algorithm and the optimization procedure. We see during convergence that the reconstructions loss is a little higher if we hadn't introduced adversarial and discriminative losses. Here, we can expect a trade-off between finding components with optimal reconstruction, adversarial and discriminative loss that depends on $\gamma_1$ and $\gamma_2$. The optimal value of these hyperparameters is selected using cross-validation, which we explain in the Section \ref{sec:robust_c_exp}. 

\section{Materials}
\label{sec:robust_real_dataset}
\subsection{Participants}


Five worldwide study samples totaling $1657$ participants, including $733$ with MDD and $924$ healthy  controls (HC) contributed T1-weighted structural scans and resting-state fMRI data (rs-fMRI) to this  study. The included cohorts combine data from the following studies: EMBARC ($4$ centers across the  United States of America, \cite{trivedi2016establishing}), University of Oxford (United Kingdom, \cite{godlewska2014short,godlewska2018predicting}), Sichuan University  
Cohort (China, \cite{zhao2020aberrant}) and STRADL (United Kingdom, \cite{navrady2018cohort,stolicyn2020automated}). Patient and controls were on average $68.6\%$ ($50- 73.8\%$) and $54.5\%$ ($53-70\%$) female. The mean age across samples was $47$ ($18-78$) years in patients and  $61$ ($16-84$) years for controls. All patients in EMBARC, Oxford and SCU were medication-free, and $15$ in SNAP and $170$ in STRADL were medicated at the  time of scanning and had a primary diagnosis of MDD that was a first episode or recurrent. MDD diagnosis was based on standardized diagnostic criteria: DSM-IV (Oxford) and  DSM-IV-TR (EMBARC, Stanford, STRADL, SCU) \cite{frances1995dsm,first2004dsm}. Table \ref{tbl:full_summary} summarizes the number of healthy and MDD participants in each site with their age and sex distribution. 

\begin{table}
\small

\centering
\begin{tabular}{@{} l*{9}{>{$}c<{$}} @{}}
\toprule
Site & \multicolumn{3}{c@{}}{$N_{\textnormal{HC}}$}& \multicolumn{3}{c@{}}{$N_{\textnormal{MDD}}$}\\
\cmidrule(l){2-7}
& \text{Number} & \% \text{of F} & \text{Age (y)} & \text{Number} & \% \text{of F} & \text{Age (y)} \\
\midrule
EMBARC-CU & 12 & 60.0 & [18,54](34) & 77 & 68.3 & [18,64](30) \\
EMBARC-MG  & 10 & 66.6 & [18,65](28) & 52 & 56.2 & [18,64](28.5) \\
EMBARC-TX & 11 & 50.0 & [23,57](26) & 97 & 67.7 & [19,65](44) \\
 EMBARC-UM & 10 & 70.0 & [23,62](41.4) & 59 & 71.1 & [18,65](31)\\
Oxford & 31 & 58.0 & [19,58](28) & 39 & 60.5 & [20,61](27)\\
SCU & 40 & 55.0 & [16,57](26.5) & 30 & 50.0 & [18,60](30.5)\\
SNAP & 55 & 65.3 & [19,58](28.8) & 55 & 63.6 & [20,56](28.2)\\
STRADL  & 755 & 53.1 & [26,84](62) & 324 & 73.8 & [26,78](60)\\ 
\midrule
Total & 924 & 54.5 & [16,84](61) & 733 & 68.6 & [18,78](47) \\
\bottomrule
\end{tabular}
\caption{Demographic characteristics of participants.}
\label{tbl:full_summary}
\end{table}


\subsection{Data Preprocessing}
\label{sec:pre}
We used FMRIB Software \cite{jenkinson2012fsl} and applied UK Biobank pipeline as a part of initial pre-processing. The data pre-processing is the same as in \citet{sahoo2021learning}. The pre-processing included the removal of the first five volumes, FSL's MCFLIRT \cite{jenkinson2012fsl} for head movement correction, global 4D mean intensity normalization, and temporal high-pass filtering ($>0.01$ Hz). To remove structured artifacts, we used FMRIB’s ICA-based Xnoiseifier FIX \cite{salimi2014automatic,griffanti2014ica}. FLIRT was then used to co-register functional images to the T1 image with Boundary Based Registration (BBR) as the cost function and FSL's FNIRT (non-linear registration) was used to register T1-weighted images to the MNI152 template. We used Independent Components Analysis (ICA) maps \cite{smith2014group} as our parcellation scheme to project the data into a $100$ dimensional space i.e., each subject had $100$ nodes after applying the parcellation scheme. Group Information Guided ICA (GIGICA) is used to extract one time series per node by mapping the group independent component spatial maps onto each subject's time series \cite{du2013group}.

\section{Experiment}
\label{sec:robust_c_exp}

\subsection{Evaluating predictive performance}
We use two different strategies to evaluate the performance of dis-rshSCP. In the first strategy, we compare five-fold cross-validation accuracy. The goal here is to check how well our model is able to estimate $\mathbf{\Lambda}$ to classify MDD vs. healthy people. We train the model on $80\%$ training set for each fold and test on the remaining. Here training is referred to solving equation \ref{eq:joint_mdd} to estimate $\mathcal{W},\mathcal{D},\mathcal{U},\mathcal{Z},\mathbf{V}$ and $\mathbf{b}$. During the test, we fix $\mathbf{b}$, estimate the rest of the parameters and evaluate the performance of the estimated parameters using $\mathbf{b}$ estimated from training. We use stratification to ensure each fold is representative of all strata (age, sex, site, and MDD) of the data. This is performed to ensure each class is approximately equally represented across each test and training fold. In the second strategy, we compare leave one site out accuracy. We train the model on all sites except one and test the model on the data from the remaining site. We compare the performance of 4 different versions of the model: 1) vanilla hSCP without reduction of any covariates, 2) dem-rshCP model with reduction of effects of age and sex, 3) site-rshSCP model with a reduction in site effects and 4) com-rshSCP model with a reduction in age, sex and site effects. We fix $k_2 = 4$ for better interpretability and based on previous experiments of \citet{sahoo2020hierarchical} and \citet{sahoo2021learning}, and find optimal value of $k_1$ from the set $\{5,10,15,20 \}$. Note that even if we select a large value of $k_2$, only a few of those components will be used to predict MDD and our experiments show that it is less than $4$. Optimal value of hyperparameters $\mu$, $\gamma_1$, $\gamma_2$ and $\lambda_1$ are selected from $[0.1,1,5]$, $[.5,1,5]$, $[0.1,1,5]$ and $[0.1,1,10]$, respectively.  
\begin{table}[t!]
\centering
        \begin{tabular}{ p{2.5cm}cccc }
 \toprule
 {Method} & {$k_1=5$} & {$k_1=10$}  & {$k_1=15$} & {$k_1=20$} \\
 \midrule
hSCP & $0.569\pm0.024$    & $0.571 \pm 0.022$ &   $0.576 \pm 0.022$ & $0.578 \pm 0.019$\\
dem-rshSCP & $0.618\pm0.021$    & $0.626 \pm 0.019$ &   $0.635 \pm 0.014$ & $0.638 \pm 0.015$\\
site-rshSCP  & $0.667 \pm 0.016$     & $0.672 \pm 0.020$ &  $0.673 \pm 0.014$  & $0.687 \pm 0.015 $ \\
comp-rshSCP & $\mathbf{0.703 \pm 0.018}$     & $\mathbf{0.727\pm 0.015}$ &  $\mathbf{0.728\pm 0.016}$  & $\mathbf{0.731\pm 0.013}$ \\
 \bottomrule
\end{tabular}
\caption{Five fold cross validation for $k_2 = 4$ (mean $\pm$ standard deviation). }
\label{tbl:five_fold_1}
\end{table}
\begin{table}[t!]
\centering
        \begin{tabular}{ p{2.5cm}cccc }
 \toprule
 {Method} & {$k_1=5$} & {$k_1=10$}  & {$k_1=15$} & {$k_1=20$} \\
 \midrule
hSCP & $0.579\pm0.036$    & $0.583 \pm 0.029$ &   $0.585 \pm 0.014$ & $0.593 \pm 0.024$\\
dem-rshSCP & $0.618\pm0.038$    & $0.626 \pm 0.027$ &   $0.635 \pm 0.036$ & $0.638 \pm 0.032$\\
site-rshSCP  & $0.620 \pm 0.037$     & $0.632 \pm 0.050$ &  $0.634 \pm 0.038$  & $0.648 \pm 0.026 $ \\
comp-rshSCP & $\mathbf{0.651 \pm 0.031}$     & $\mathbf{0.680\pm 0.024}$ &  $\mathbf{0.681\pm 0.042}$  & $\mathbf{0.689\pm 0.047}$ \\
 \bottomrule
\end{tabular}
 \caption{Leave one site accuracy for $k_2 = 4$ (mean $\pm$ standard deviation).}
  \label{tbl:leave_one_1}

\end{table}

Table \ref{tbl:five_fold_1} and Table \ref{tbl:leave_one_1} show results of five-fold and leave one site cross-validation. From the results, we can see the baseline method's dull prediction performance, which shows the difficulty of the challenge we are tackling. It can be seen that dis-rshSCP has a better performance than vanilla hSCP. Out of demographics and site as a covariate, we see that reducing site variability has a major impact on the prediction performance compared to reducing demographics information. The best performance is achieved when we remove both demographics and site information which suggests that removing heterogeneity from the data can help improve the predictability power of these components, thus giving more reliable components discriminative of MDD.

\subsection{Reproducibility}
We have shown that our method can extract components with high predictability power. This section shows that these components are highly reproducible and generalizable, which is important for the future application of our framework. We use split-sample reproducibly to measure the generalizability of the components, which measures how likely a set of components is replicable across the same population. First the optimal parameters are selected based on the highest five fold cross validation, then the split sample reproducibility is computed by dividing the dataset into two random splits with the same stratification and calculating the correlation between components derived from the two splits. 
Table \ref{tbl:split_sample_1} shows split sample reproducibility of the components extracted from the different methods. In all the experiment, results are generated by computing reproducibility over $20$ runs. It can be seen that the results are similar to prediction performance results, i.e., the components are highly reproducible, and reducing demographics and site information helps in improving the reproducibility. This suggests even after adding prediction loss, the method can find highly reproducible components.
\begin{table}[t!]
\centering
        \begin{tabular}{ p{2.5cm}cccc }
 \toprule
 {Method} & {$k_1=5$} & {$k_1=10$}  & {$k_1=15$} & {$k_1=20$} \\
 \midrule
hSCP & $0.752\pm0.032$    & $0.713 \pm 0.037$ &   $0.688 \pm 0.039$ & $0.621 \pm 0.021$\\
dem-rshSCP & $0.781\pm0.024$    & $0.749 \pm 0.045$ &   $0.739 \pm 0.038$ & $0.688 \pm 0.032$\\
site-rshSCP  & $0.782 \pm 0.031$     & $0.739 \pm 0.032$ &  $0.728 \pm 0.025$  & $0.661 \pm 0.028 $ \\
comp-rshSCP & $\mathbf{0.805 \pm 0.029}$     & $\mathbf{0.761\pm 0.026}$ &  $\mathbf{0.725\pm 0.035}$  & $\mathbf{0.692\pm 0.031}$ \\
 \bottomrule
\end{tabular}
\caption{Split sample reproducibility for $k_2 = 4$ (mean $\pm$ standard deviation). }
\label{tbl:split_sample_1}
\end{table} 
 \begin{figure}[t!]
  \centering
    \begin{minipage}{.49\linewidth}
  \centering
   \subcaptionbox{}
   {\makebox[0.97\linewidth][c]{\includegraphics[trim={1cm 1cm 1cm 1cm},clip,width=.60\linewidth]
    {Figures/somatomotor_default_mode_ventral_attention.png}}}
        \end{minipage}
\begin{minipage}{.49\linewidth}
\centering
 \subcaptionbox{}
 {\makebox[0.97\linewidth][c]{\includegraphics[trim={1cm 1cm 1cm 1cm},clip,width=.60\linewidth]
    {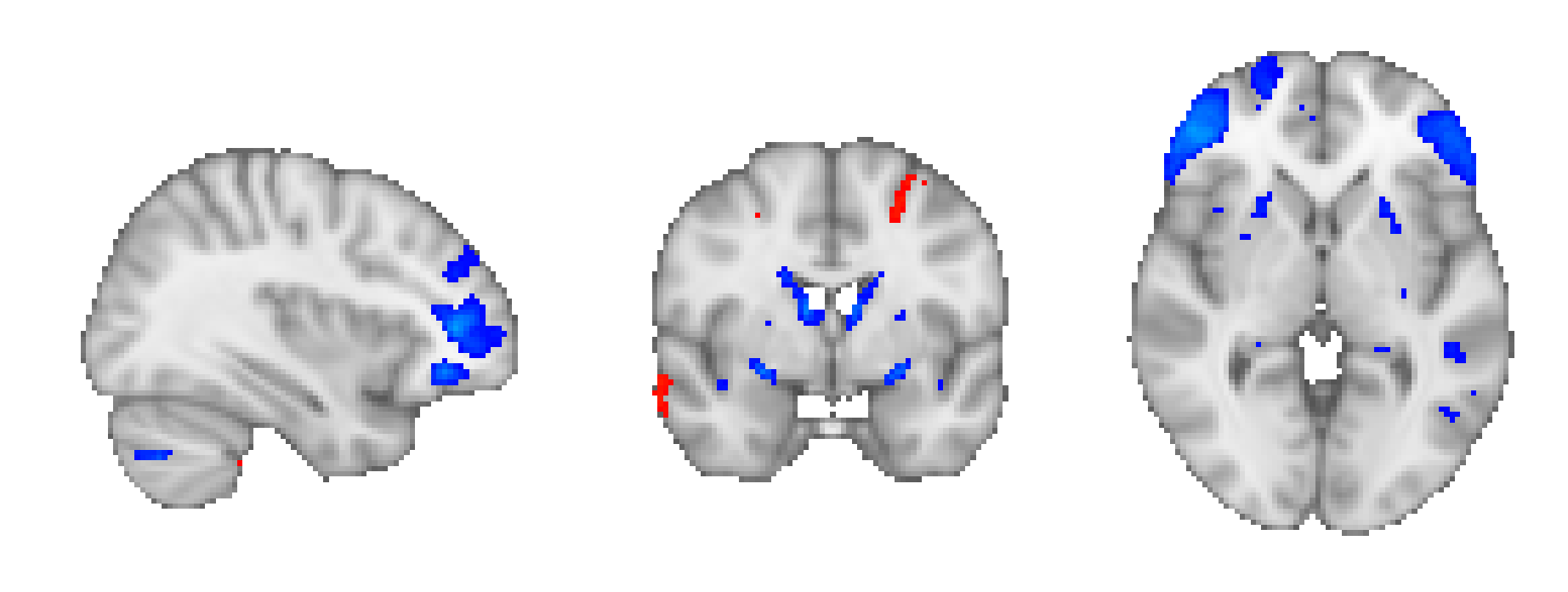}}}
  \end{minipage}
  
     \begin{minipage}{.32\linewidth}
  \centering
   \subcaptionbox{}
   {\makebox[0.97\linewidth][c]{\includegraphics[trim={1cm 1cm 1cm 1cm},clip,width=.85\linewidth]
    {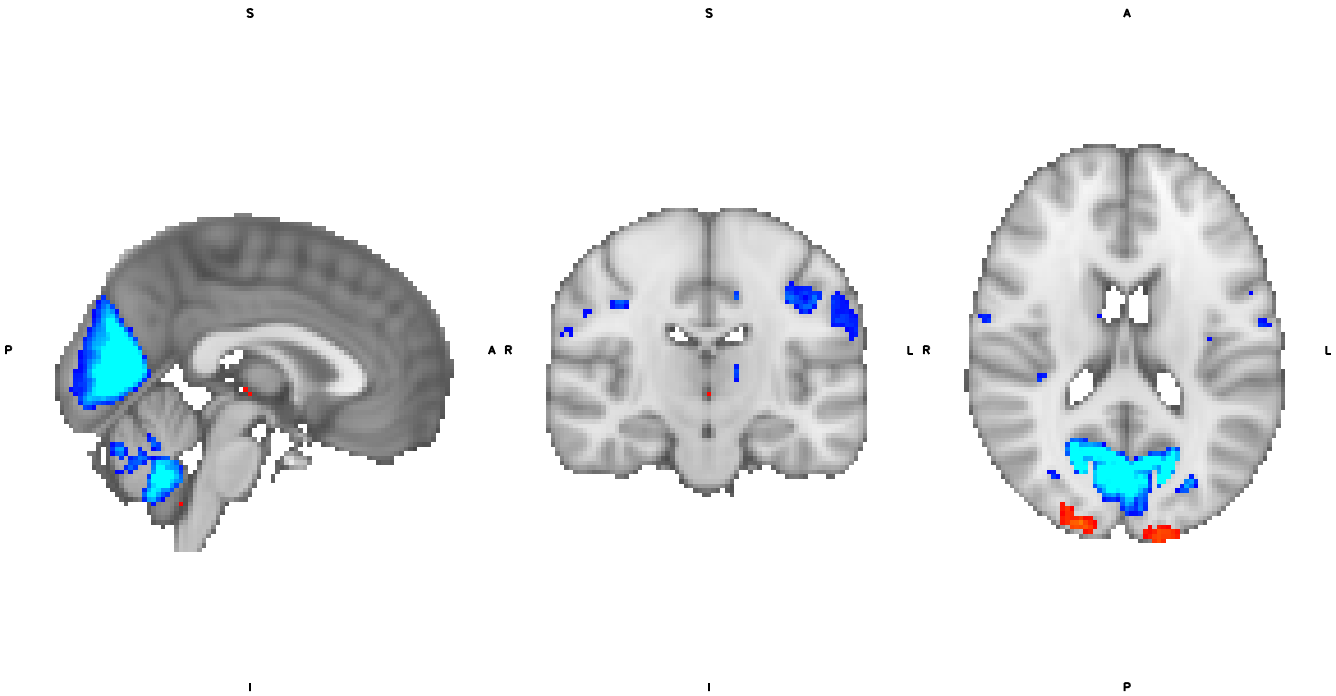}}}
        \end{minipage}
\begin{minipage}{.32\linewidth}
\centering
 \subcaptionbox{}
 {\makebox[0.97\linewidth][c]{\includegraphics[trim={1cm 1cm 1cm 1cm},clip,width=.85\linewidth]
    {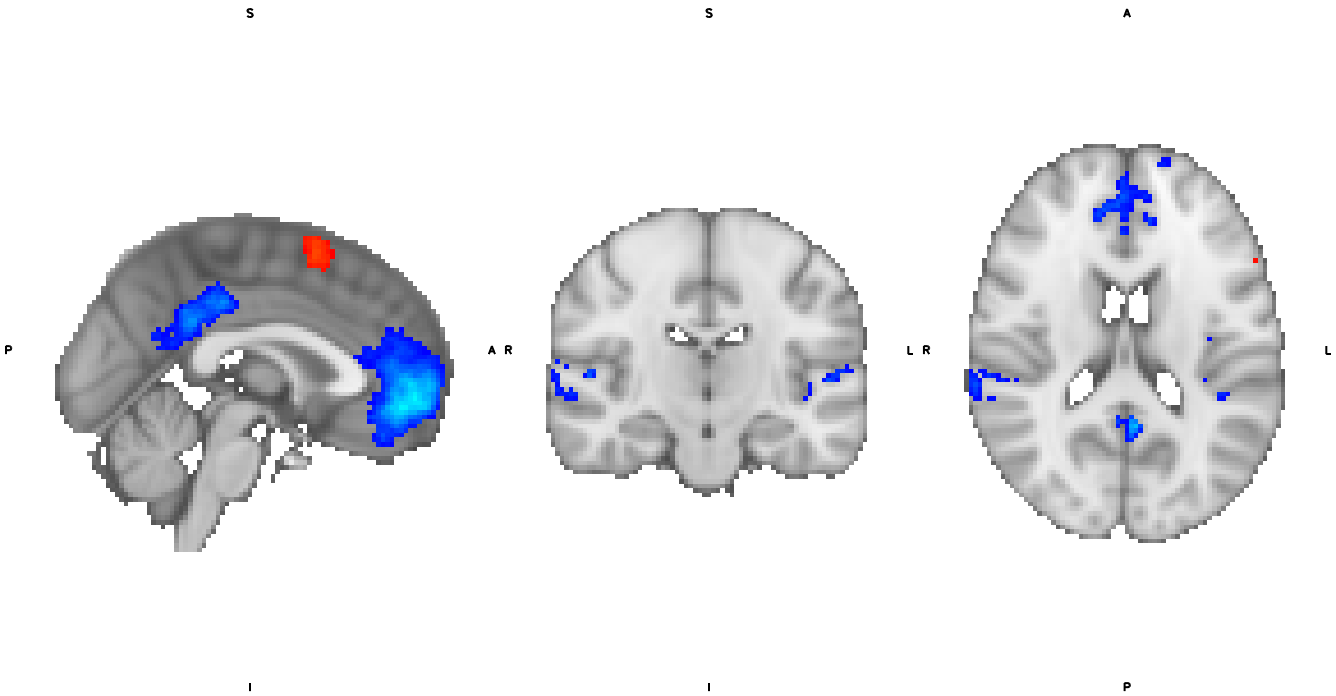}}}
  \end{minipage}
  \begin{minipage}{.32\linewidth}
\centering
 \subcaptionbox{}
 {\makebox[0.97\linewidth][c]{\includegraphics[trim={1cm 1cm 1cm 1cm},clip,width=.85\linewidth]
    {Figures/Images1/visual_CEN}}}
  \end{minipage}
\caption{(1) SMN, DMN and SN, (2) DMN and CEN, (3) VN, (4) DMN and (5) VN and CEN. We use red and blue colors to represent two different sub-networks which are anti-correlated with each other and regions with a colored part is correlated among themselves. \label{fig:relation1}  }
\end{figure}
\subsection{Analysis of Components}
We select $k_1$ based on optimal tradeoff between classification accuracy and reproducibility, the result of which are provided in Table \ref{tbl:five_fold_1}, \ref{tbl:leave_one_1} and \ref{tbl:split_sample_1}. We can see from the Table \ref{tbl:five_fold_1} and \ref{tbl:leave_one_1} that classification accuracy starts to plateau at $k_1 = 10$, but if we look at reproducibility, it continues to decrease linearly. Thus we choose $k_1$ to have good classification accuracy without losing reproducibility of the components. Our aim here is to analyze components that are most predictive of MDD. For this, the components are selected on basis of $\mathbf{b}$ in the logistic regression (\ref{eq:logistic}). We perform a hypothesis test with the null hypothesis being that the $k$th component is not discriminative of MDD, i.e., ${b}_k$ is $0$ and $\text{p-value} < 0.05$. If a component has a significant positive value of ${b}_k$, then it is likely to have a higher weight in the MDD population than the healthy population and vice versa. Each column of $\mathbf{W}$ contains a component storing information about a set of co-activated regions which can be positively or negatively correlated with each other, and we map normalized values of each column onto corresponding regions. On the basis of hypothesis test, we obtained $5$ significant components out of $10$. 

\begin{figure}[t]
\centering
\begin{forest}
  styleA/.style={top color=white, bottom color=white},
  styleB/.style={%
    top color=white,
    bottom color=red!20,
    delay={%
      content/.wrap value={##1\\{\includegraphics[scale=.5]{hand}}}
    }
  },
for tree={
    edge path={\noexpand\path[\forestoption{edge}] (\forestOve{\forestove{@parent}}{name}.parent anchor) -- +(0,-16pt)-| (\forestove{name}.child anchor)\forestoption{edge label};}
},
  forked edges, s sep=20mm
[{ \includegraphics[trim={1cm 1cm 1cm 1cm},clip,scale=0.11]{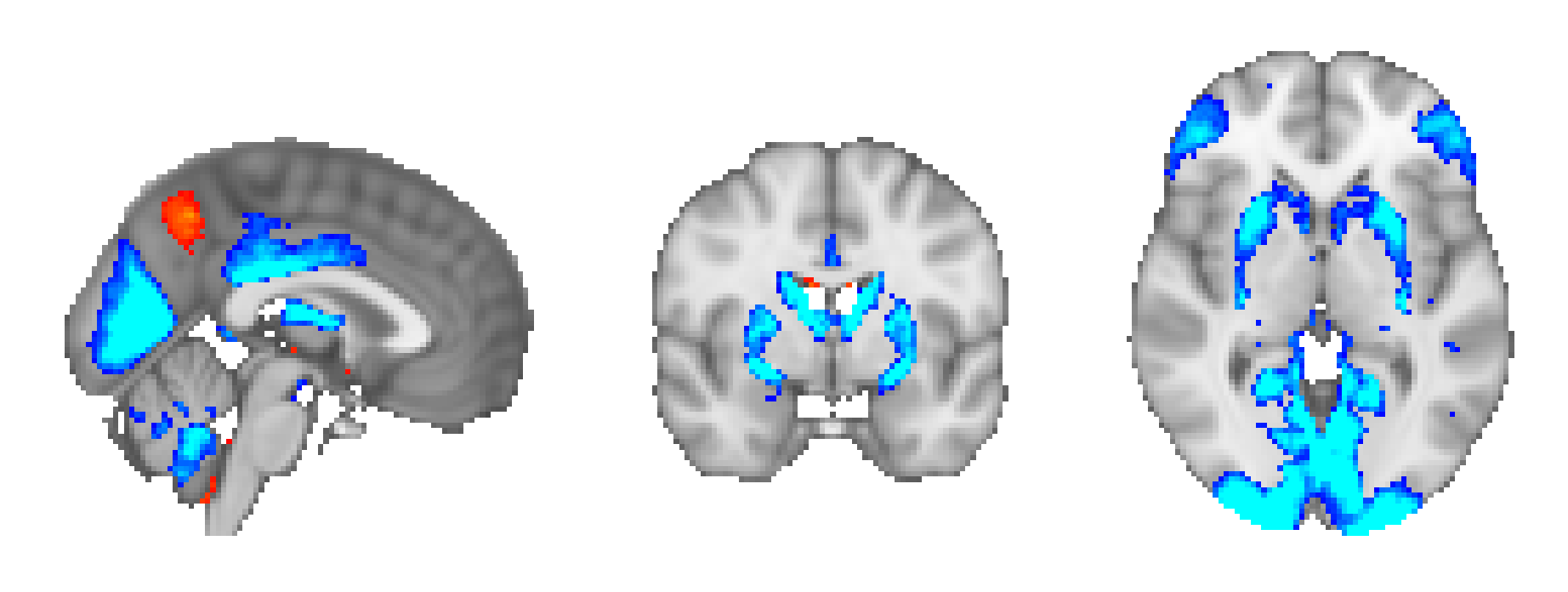}}
    [5{ \includegraphics[trim={0cm 1cm 1cm 0cm},clip,scale=0.11]{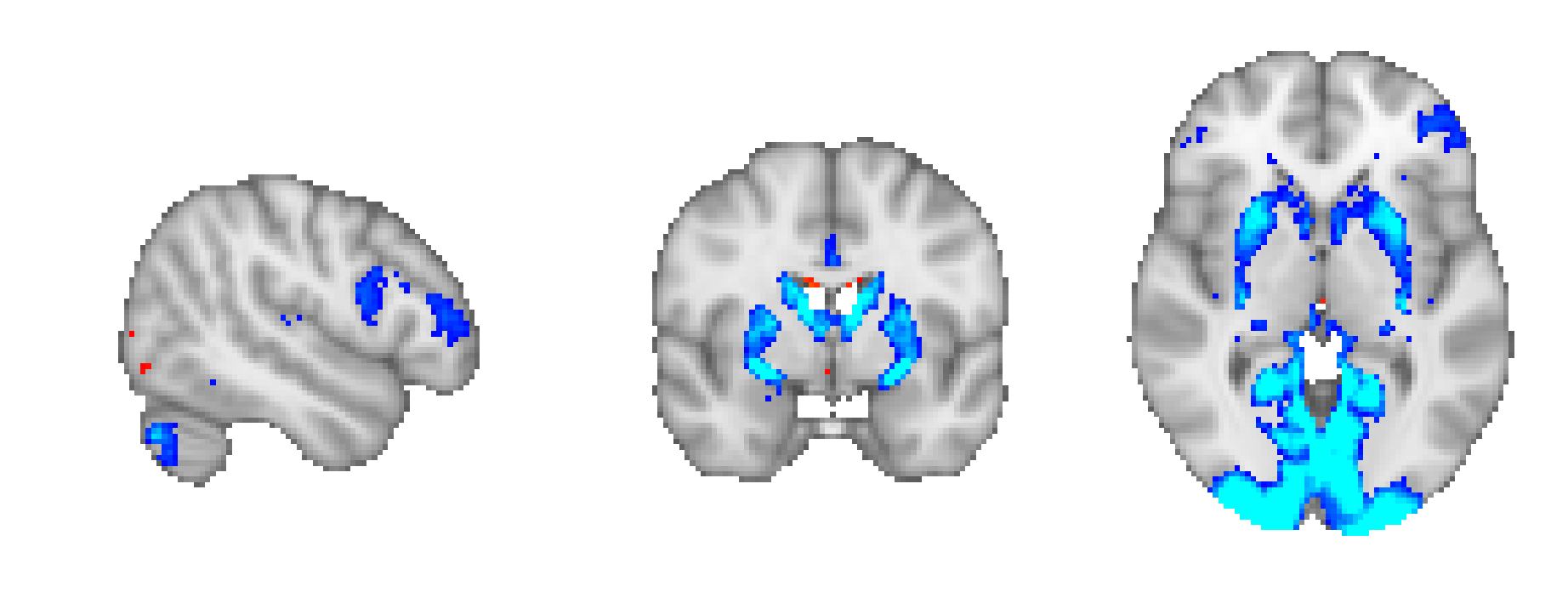}}]
    [2{ \includegraphics[trim={0cm 1cm 1cm 0cm},clip,scale=0.11]{Figures/Images1/DMN_CEN.png}}]
     ]
  ]
\end{forest}
\caption{Hierarchical component having significant predictive power. Here the coarse-scale component is comprised of component $5$ and $2$ consisting of regions of DMN, CEN, and VN. \label{fig:hirr}}
\end{figure}

\begin{figure}
    \begin{minipage}[t]{.47\textwidth}
        \centering
        \includegraphics[width=\textwidth]{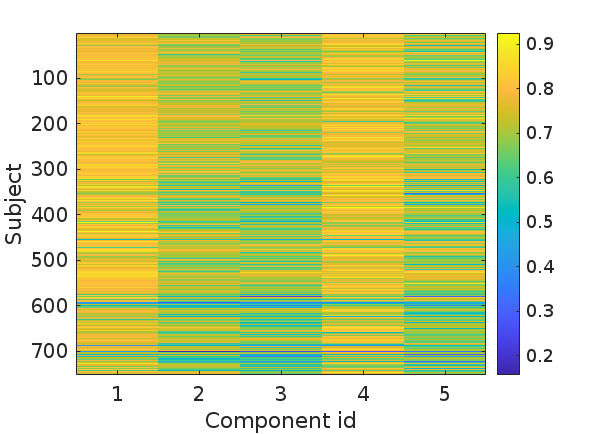}
        \subcaption{Heterogeneity captured by MDD subjects}
    \end{minipage}
    \begin{minipage}[t]{.47\textwidth}
        \centering
        \includegraphics[width=\textwidth]{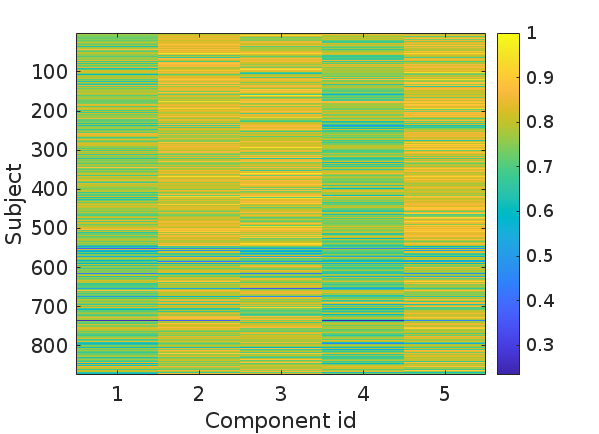}
        \subcaption{Heterogeneity captured by healthy subjects}
    \end{minipage}

        \caption{Heterogeneity captured by components $1-5$. The color represents the strength ($\mathbf{\Lambda}_1$) of each component in each individual. We normalized $\mathbf{\Lambda}_1$ for comparison purposes with $1$ being the highest value and $0$ lowest.}
    \label{fig:hetero}
\end{figure}

Figure \ref{fig:relation1} displays components most predictive of MDD.The color bar indicates the contribution of each region in the components. Regions colored blue are anti-correlated with displayed areas in red. The extracted components store information between various parts of the human brain, i.e., whether the regions are correlated or anti-correlated with each other. These regions could be smaller parts of the brain or clustered to form resting-state functional networks. Component $1$ comprises of regions of Somatomotor Network (SMN), Default Mode Network (DMN) and Salience Network (SN), where regions of DMN and SMN are anti-correlated with SN. Component $2$ comprises of regions of DMN and Central Executive Network (CEN). Component $3$ and $4$ consists of regions of DMN and Visual Network (VN), and component $5$ consists of regions of VN and CEN. In addition to the $5$ fine scale significant components, we also recovered a significant hierarchical component comprising of $5$ and $2$ shown in Figure \ref{fig:hirr}. The strength of each component in MDD and healthy individuals is shown in Figure \ref{fig:hetero}. We observe decreased representation  components $2$, $3$ and $5$ comprising the DMN, CEN and VN in MDD subjects compared to healthy. Increased representation is seen in the components comprising SMN, DMN and SN in MDD subjects compared to healthy.

\section{Discussion}
\label{sec:robust_disc}

Our aim was to identify sparse hierarchical connectivity patterns discriminative of MDD and we achieved this with three coupled loss functions: 1) representation learning loss, 2) discriminative loss, and 3) adversarial loss. Our model cleverly exploits the rs-fMRI correlation matrix structure to extract sparse patterns through the representation learning loss. The model also serves as a dimensionality reduction technique helping to extract low-rank sparse decomposition. The classification loss in the model forces the decomposition to extract MDD specific group-level patterns. Notice that there is a slight tradeoff in classification performance at the expense of representation learning loss. We highlight this as it is essential for exploration. 

This work provides proof-of-principle analysis focusing on age, sex, and site as diversity factors in the dataset. Adversarial loss helps reduce these factors, resulting in improved reproducibility, generalizability and prediction performance. Our results show that the largest improvement is acheived when accounting for all factors instead of focusing on just one. However, other factors could be considered for future work, such as comorbidity, ethnicity/race, open vs. closed eye during fMRI acquisition, etc. These factors could help further improve pattern stability and classification performance. 

Here, we used split-sample reproducibility along with cross-validation accuracy to extract components with the aim to not just have high predictive power but also high reproducibility. We note here that after a certain number of components, the addition of more components does not increase the accuracy. However, the addition of components decreases the reproducibility, which can be attributed to the model capturing noisy components which are not predictive of MDD. Future investigations could look at various other reproducibility measures and classification metrics, and analyze optimality of components in multiple settings to assess the impact on downstream analysis. 

We identified $5$ components that are highly predictive of MDD status. These components store information about the inter- and intra-connectivity within networks and suggest that MDD is characterized by disruptions in the following networks: intra-connectivity in the visual and default mode networks, inter-connectivity between the visual and central executive networks and inter-connectivity between the salience, default mode and central executive network. These discriminative patterns revealed by our framework are consistent with the recent literature on changes in functional connectivity patterns in MDD.  

Impaired visual perception has been found in patients with MDD. It is considered an important aspect of the disease, whereby there is a positive correlation between the degree of visual disruption and the severity of symptoms \cite{song2021reduction} and reduced visual network connectivity \cite{zeng2012identifying,veer2010whole}.
Studies suggest that perceptual impairments are linked to abnormal cortical processing and disrupted neurotransmitter systems whilst retinal processing remains intact \cite{nikolaus2012key,salmela2021reduced}. 


The DMN has frequently been implicated in MDD pathophysiology due to its role in producing negative, self-referential, ruminative thoughts \cite{hamilton2015depressive}. There are previous reports of both hypo- and hyper-connectivity within the DMN in depression \cite{kaiser2015large,tozzi2021reduced} but \citet{liang2020biotypes} suggest that depression is characterized by two subgroups of patients exhibiting opposing dysfunctional DMN connectivity. These inconsistent findings could be due to sample variations in symptom profiles as variability in connectivity within the DMN is positively correlated with levels of ruminative thoughts \cite{wise2017instability} while hypo-connectivity has been associated with symptom severity in recurrent MDD \cite{yan2019reduced}. 

The inconsistent abnormalities of DMN connectivity in depression suggest that it could instead be the interplay between the DMN and other networks that leads to the variety of symptoms observed in depression. Indeed, our patient sample highly expressed abnormal connectivity across three networks, the CEN, DMN and SN, which have been put forward as being part of a triple network model of psychopathology \cite{menon2011large}. In this model, aberrant saliency attribution within the SN weakens the engagement of the CEN and disengagement of the DMN, leading to cognitive and emotional problems. Therefore, identifying hierarchical connectivity patterns is crucial to understanding the interaction between networks that give rise to disease. These findings show that the proposed method can extract meaningful components with high reproducibility and clinical relevance without traditional seed-based methods, which rely too heavily on a priori regions of interest. In a nutshell, these findings could further our understanding of MDD from a functional network perspective.

There are several future directions from methodological and clinical perspectives. First, the model robustness can be improved by introducing masking of correlation matrices. It has been shown that masking of features \cite{devlin2019bert} while learning can improve the robustness of the model and its predictability power. Second, instead of reducing age and sex related heterogeneity, one could disentangle components and learn age, sex and MDD specific components. Our method is limited to finding effects of MDD on brain connectivity; in the future, we will use the proposed approach to study the effects of antidepressant medications on brain connectivity in MDD in resting-state or task-based fMRI \cite{gudayol2015changes,brakowski2017resting}. Another important direction would be to combine structural connectivity information using Diffusion Tensor Imaging (DTI) in our optimization model. Unifying structural, functional and disease information would give a more comprehensive view of neurobiological abnormalities and altered brain functioning and improve MDD diagnosing ability.

Our proposed model has a few weaknesses, adding directions for future work. First, we only consider logistic regression as our classification model, the results of which might be sub-optimal. Our method can be modified to include various classifiers, and a comparison study can be performed to find the optimal classifier with more emphasis on the classification performances. Here, using SVM like optimization models might be more straightforward than incorporating XGBoost like models. In this study, our focus was on the interpretation of the components; we did not evaluate the reproducibility and classification accuracy for broad values of $k_1$ and $k_2$. Future models could benefit from a thorough investigation of this shortcoming. However, this analysis is beyond the scope of this work. Second, the robust to covariate model is a multi-task learning model whose loss weights are manually selected. More advanced techniques such as the Pareto multi task learning model \cite{lin2019pareto} and balanced multi task learning framework \cite{liang2020simple} can be used to improve the results. Our study lumps all the MDD patients together and then analyzes the change in functional networks, but previous studies have signaled that it is a highly heterogeneous psychiatric disorder \cite{hyman2008glimmer,miller2010beyond}. Clustering approaches can be used on $\mathbf{\Lambda}$ to find subtypes of MDD, the analysis of which is beyond the scope of the current paper.

\section{Conclusion}
\label{sec:conclusion}

This work presents an effective matrix decomposition strategy to combine rs-fMRI data with clinical information. Our framework is completely data-dependent and makes minimal assumptions about the data. We extended the method of reducing site effects in hSCP by adding additional loss terms for reducing age and sex effects to estimate robust components discriminative of MDD. We added a discriminator to extract components that represent the MDD population. The problem is formulated as a minimax non-convex optimization problem and is solved using adaptive gradient descent. Experimentally, using a pooled dataset from five different sites, we showed that reducing heterogeneity introduced by age, sex, and site could improve the prediction capability of the components, which is validated using fivefold and leave one site out cross-validation. Our framework robustly identiﬁes brain patterns characterizing MDD and provides an understanding of the manifestation of the disorder from a functional networks perspective. Moreover, our evaluation on a large multi-site dataset validates the reproducibility and generalizability of the framework. In addition, our model is not limited to MDD and can be easily adapted to other disorders such as ASD, ADHD, etc. Moreover, it can easily incorporate other models outside the medical domain, provided we have access to valid network measures as an input. This greatly broadens the method's applicability to numerous applications from varied fields.

\footnotesize
\bibliographystyle{kbib}
\bibliography{main.bib}

\normalsize

\newpage
\appendix

\section{Related Work: Extended}
\label{sec:rel_work}

Several different methods have been developed to analyze brain organization using fMRI data. One of the widely used methods include Independent Component Analysis \cite{smith2009correspondence}, Non-Negative Matrix Factorization \cite{potluru2008group},Sparse Dictionary Learning \cite{lee2010data,eavani2015identifying} and matrix factorization based Granger Causality \cite{sahoo2018gpu,sahoo2019sparse}. These methods are used to estimate interpretable patterns capturing functional human brain information. Another commonly used method is multi-scale community detection methods to understand the hierarchical organization of the human brain \cite{ferrarini2009hierarchical,al2018tensor,akiki2019determining,ashourvan2019multi}. The approaches mentioned above have one or more limitations: 1) the assumption that the components are independent, 2) the inability to capture a subject-specific representation of the patterns, and 3) removal of negative edge links in the method because the negative links are treated as repulsions. Recently, several deep learning based methods \cite{zhang2020hierarchical,huang2017modeling,hu2018latent,zhang2020hierarchical,dong2019modeling} have been introduced to estimate functional networks. These methods have shown promising results in terms of prediction performances. Still, there are one or more disadvantages: 1) not capturing negative correlation between nodes in a pattern resulting in the lacking inference of inter-network connectivity, 2) the inability to capture a subject-specific representation of the patterns, and 3) ``black-box'' results due to non-linearity of the deep learning model causing loss of interpretability.

Hierarchical Sparse Connectivity Patterns (hSCPs) \cite{sahoo2021learning} is a dimensionality reduction method overcoming the disadvantages mentioned above. The method allows us to investigate hierarchically organized co-activated brain regions and probe individual representations of brain activity patterns. We consider hSCP as a method that produces a small set of components consisting of functionally synchronous brain regions that statistically appear to have relatively more synchronous activation patterns and hence could be presumed to be parts of underlying brain networks. These components can capture interactions between various resting-state functional connectivity networks, for example, Default Mode Network, Salience Network, etc. However, these networks need not necessarily be present in each individual or subset of individuals.

\section{Algorithm}
\label{sec:alg}
\subsection{Alternating Minimization}
We employ the same alternating minimization technique as used in previous hSCP papers \cite{sahoo2021learning,sahoo2021extraction,sahoo2020hierarchical} for estimating model parameters. Here, we optimize the objective function \ref{eq:joint_mdd} for each variable using adaptive gradient descent (AMSGrad) \cite{reddi2019convergence} while holding estimates of other variables as constants. $\beta_1$ and $\beta_2$ value in AMSGrad are kept to be $0.9$ and $0.999$. The gradients of each variable  used in gradient descent is defined in the next section. Alternating minimization procedure to solve equation \ref{eq:joint_mdd} is described in Algorithm~\ref{alg:site_HSCP}. The algorithm can be modified for solving equation \ref{eq:test} by commenting out the gradient descent of $\mathbf{b}$. $\mathcal{U}$ and $\mathbf{V}$ are initialized using equation $10$ in \cite{sahoo2021learning} and $\svdinit$ algorithm \cite{sahoo2020hierarchical}[Algorithm 2] is used to initialize $\mathcal{W}$ and $\mathcal{D}$. Additional constraints in the optimization problem are met by using below operators:
\begin{itemize}[leftmargin=5.5mm]
    \item $\proj_1(\mathbf{W},\lambda)$- Projecting each column of $\mathbf{W}$ into the intersection of $L_1$ and $L_\infty$ ball
    \item $\proj_2$- Projecting a matrix onto $\mathbb{R}_{+}$ space
    \item $\proj_3$- Projecting a vector onto $L_1$ ball
\end{itemize}
 \begin{algorithm}[t]
 \caption{dis-rshSCP}\label{alg:site_HSCP}
\begin{algorithmic}[1]
\State \textbf{Input:} Data $\mathcal{C}$, number of connectivity patterns $k_1$, \ldots, $k_K$ and sparsity $\lambda_1$, \ldots, $\lambda_K$ at different level, hyperparameters $\mu$, $\gamma_1$ and $\gamma_2$. 
\State Initialize $\mathcal{W}$ and $\mathcal{D}$ using $\svdinit$
\State Initialize $\mathcal{U}$ and $\mathbf{V}$ using $\siteinit$
\Repeat 
\For{$r=1$ {\bfseries to} $K$} 
\If{Starting criterion is met}
\State $\zeta \leftarrow \descent(\zeta)$
\State $\mathbf{b} \leftarrow \descent(\mathbf{b})$
\EndIf
\If{$r==1$}
   \State $\mathbf{W}_r \leftarrow \proj_1(\mathbf{W}_r,\lambda_r) $
   \Else{}
\State $\mathbf{W}_r \leftarrow \proj_2(\mathbf{W}_r) $
   \EndIf
\For{$n= 1,..,N$} 
\State $\mathbf{\Lambda}_r^n \leftarrow \descent(\mathbf{\Lambda}_r^n)$
\State $\mathbf{\Lambda}_r^n \leftarrow \proj_2(\mathbf{\Lambda}_r^n) $
\EndFor
\For{$s=1$ {\bfseries to} $S$}
\State $\mathbf{U}^s_r \leftarrow \descent(\mathbf{U}^s_r)$
\EndFor
\State $\mathbf{V}_r \leftarrow \descent(\mathbf{V}_r)$
\State $\mathbf{V}_r \leftarrow \proj_3(\mathbf{V}_r,\mu)$
\EndFor
\Until{Stopping criterion is reached}
\State \textbf{Output:} $\mathcal{W}$, $\mathcal{L}$ and $\mathbf{b}$
\end{algorithmic}
\end{algorithm} 
\subsection{Gradient Calculations}
This section defines the gradients used in alternating gradient descent algorithm, some of which are already defined in \citet{sahoo2021learning}. Let
\begin{align*}
\mathbf{\tilde{W}}_0 &= \mathbf{W}_0 = \mathbf{I}_P, \qquad
\mathbf{Y}_r = \prod_{j=0}^{r}\mathbf{W}_j, \qquad 
\mathbf{Q}_{m,n}^r = (\prod_{j=1}^{m-r}\mathbf{W}_j)\mathbf{Z}^s_{m-r}\mathbf{\Lambda}^n_{m-r}(\prod_{j=1}^{m-r}\mathbf{W}_j)^\top, \\
\mathbf{X}^n_r &= \mathbf{\Theta}^n - \mathbf{U}^s_r \mathbf{V}_r, \qquad \mathbf{H}^n_r = \mathbf{\Theta}^n - (\prod_{j=1}^{r}\mathbf{W}_j)\mathbf{Z}^s_{m-r}\mathbf{\Lambda}_r^n(\prod_{j=1}^{r}\mathbf{W}_n)^\top, 
\end{align*}
where $n \in \mathcal{I}_s$. Let $J$ be the objective function defined in \ref{eq:joint_mdd} and then gradient of $J$ with respect to $\mathbf{\Lambda}_r^n$ is:
\begin{align*}
\frac{\partial J}{\partial \mathbf{\Lambda}_r^n} & = \frac{\partial G(\mathcal{C},\mathcal{W},\mathcal{D},\mathcal{U},\mathbf{V},\mathcal{Z})}{\partial \mathbf{\Lambda}_r^n}  - \gamma_1 \frac{\partial \mathcal{L}_{1}(\zeta, \mathcal{D}, \mathbf{y}_{site},\mathbf{y}_{age},\mathbf{y}_{sex})}{\partial \mathbf{\Lambda}_r^n} + \gamma_2 \frac{\partial \mathcal{L}_{2}(\mathbf{b}, \mathcal{D}, \mathbf{y}_{mdd})}{\partial \mathbf{\Lambda}_r^n} 
\\ &  = (-2\mathbf{Y}_r^\top\mathbf{X}_r^n\mathbf{Y}_r + 2\mathbf{Y}_r^\top\mathbf{Y}_r\mathbf{Z}^s_r\mathbf{\Lambda}_r^n\mathbf{Y}_r^\top\mathbf{Y}_r) \circ \mathbf{Z}_{r}^s +  \mathbf{F}.
\end{align*}
where is $\mathbf{F}$ is calculated using automatic differentiation tool provided by \textsc{matlab}. The gradient of $J$ with respect to $\mathbf{W}_r$ is:
\begin{align*}
\frac{\partial J}{\partial \mathbf{W}_r} &= \frac{\partial G(\mathcal{C},\mathcal{W},\mathcal{D},\mathcal{U},\mathbf{V},\mathcal{Z})}{\partial \mathbf{W}_r}\\ &= \sum_{n=1}^N\sum_{j=r}^{K} -4\mathbf{Y}_{r-1}^\top\mathbf{X}_n\mathbf{Y}_{r-1}\mathbf{W}_r\mathbf{Q}_{j,n}^r +   4\mathbf{Y}_{r-1}^\top\mathbf{Y}_{r-1}\mathbf{W}_r\mathbf{Q}_{j,n}^r\mathbf{W}_r^\top\mathbf{Y}_{r-1}^\top\mathbf{Y}_{r-1}\mathbf{W}_r\mathbf{Q}_{j,n}^r.  
\end{align*}
The gradient $J$ with respect to $\mathbf{U}^s$ and $\mathbf{V}$ is:
\begin{align*}
    \frac{\partial J}{\partial \mathbf{U}^s_r} &= \left(\sum_{n=\mathcal{I}_s} \left(\mathbf{H}^n_r - \mathbf{U}^s_r \mathbf{V}_r \right)\mathbf{V}^\top_r \right) \circ \mathbf{I}_{p} \\
    \frac{\partial J}{\partial \mathbf{V}_r} &= \sum_{s=1}^S\sum_{n \in \mathcal{I}_s} \mathbf{U}^s_r \left(\mathbf{H}^n_r - \mathbf{U}^s_r \mathbf{V}_r \right).
\end{align*}
The gradient of $J$ with respect to $\mathbf{K}_r^s$ is:
\begin{align*}
\frac{\partial J}{\partial \mathbf{K}_r^s} & = \frac{\partial G(\mathcal{C},\mathcal{W},\mathcal{D},\mathcal{U},\mathbf{V},\mathcal{Z})}{\partial \mathbf{K}_r^s} \\ &  = \sum_{n=\mathcal{I}_s}(-2\mathbf{Y}_r^\top\mathbf{X}_r^n\mathbf{Y}_r  + 2\mathbf{Y}_r^\top\mathbf{Y}_r\mathbf{Z}^s_r\mathbf{\Lambda}_r^n\mathbf{Y}_r^\top\mathbf{Y}_r) \circ \mathbf{\Lambda}_{r}^n.
\end{align*}

\subsection{Convergence results}
\begin{figure}
   \centering
\begin{subfigure}{.45\textwidth}
\centering
     \includegraphics[width=.79\linewidth]{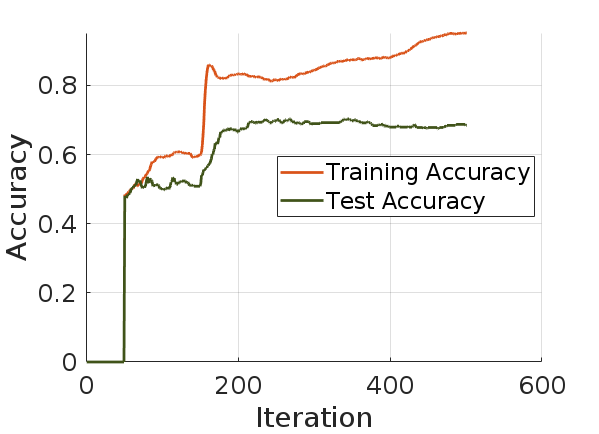}
    \caption{Training and test accuracy}   
\end{subfigure}
 \begin{subfigure}{.45\textwidth}
 \centering
           \includegraphics[width=.79\linewidth]{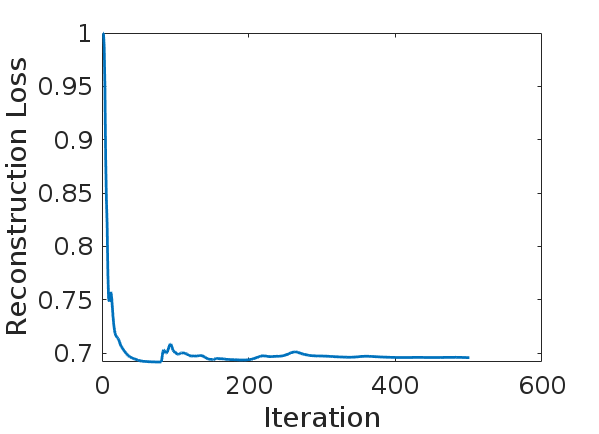}
    \caption{Reconstruction loss}  
 \end{subfigure}
    \caption{(1) Training and test accuracy and (2) reconstruction loss of dis-rshSCP algorithm for $k_1 = 10$ and $k_2 = 4$. \label{fig:convg}}
\end{figure}

Reconstruction error is used to empirically validate the convergence of the Algorithm~\ref{alg:site_HSCP}:
\begin{align*}
    \frac{ \sum_{s=1}^S \sum_{n \in \mathcal{I}_s} \sum_{r=1}^{K}||\mathbf{\Theta}^i - (\prod_{j=1}^{r}\mathbf{W}_j)\mathbf{Z}^s\mathbf{\Lambda}_r^i(\prod_{j=1}^{r}\mathbf{W}_j)^\top - \mathbf{U}^s_r \mathbf{V}_r||_F^2}{\sum_{n=1}^{N} K||\mathbf{\Theta}^i ||_F^2}.
\end{align*}
Figure \ref{fig:convg} shows the reconstruction loss, training error and test error of the algorithm on one of the folds of 5 fold cross validation. We can see from the result that the algorithm converges after $400$ iterations. The algorithm runs without the adversarial and discriminative loss for the first $50$ iterations therefore no accuracy is determined for that duration, and the reconstruction loss is constantly decreasing. As the adversarial and discriminative losses are introduced, the reconstruction loss starts to oscillate and then converges to a sub optimal loss as compared to if there were no additional losses were introduced. Whereas the training accuracy keeps on getting better but the test accuracy converges after $200$ iteration. This shows that the algorithm can overfit, hence necessary cross-validation is important for selecting the hyperparamters in the loss function \ref{eq:joint_mdd}.

\section{Multi Task Learning}
\label{sec:mtl}
  \begin{figure}
\centering
   \begin{tikzpicture}[
node distance = 5mm and 11mm,
R/.style = {fill=#1, rounded corners, 
            minimum height=2em, inner xsep=2em,
            rotate=90, anchor=center, 
            drop shadow},
N/.style = {draw, fill=#1, minimum size=2em}
                        ]
\node (r1) [R=gray]                          {Input Layer};
\node (r2) [right=22mm of r1.south,R=red!20]    {Shared Layer};
\node (n12) [N=green!30,  right=of r2.south]    {};
\node (n11) [N=blue!30,   above=of n12] {};
\node (n13) [N=yellow!20, below=of n12] {};
\node (n21) [N=blue!30,   right=of n11] {};
\node (n22) [N=green!30,  right=of n12] {};
\node (n23) [N=yellow!20, right=of n13] {};
\draw [->] (n21) -- ++ (1.1,0) node[right] {Task 1};
\draw [->] (n22) -- ++ (1.1,0) node[right] {Task 2};
\draw [->] (n23) -- ++ (1.1,0) node[right] {Task 3};
\node [align=center, above=of n11]  {Task Specific \\ Layer};
\node [align=center, above=of n21]  {Output \\ Layer};
\draw[dashed, gray] 
    (r1.south east) -- (r2.north east)
    (r1.south west) -- (r2.north west)
    ([shift={(1ex,-1ex)}] r1.south east) -- ([shift={(-.5ex,+1ex)}] r2.north west)
    ([shift={(1ex,+1ex)}] r1.south west) -- ([shift={(-.5ex,-1ex)}] r2.north east)
    ([shift={(1ex,-3ex)}] r1.south) -- ([shift={(-.5ex,+3ex)}] r2.north)
    ([shift={(1ex,+3ex)}] r1.south) -- ([shift={(-.5ex,-3ex)}] r2.north);

\draw[dashed, gray] 
     (n11.north east) -- (n21.south west)
    (n11.south east) -- (n21.north west)
    ;    
\draw[dashed, gray] 
     (n12.north east) -- (n22.south west)
    (n12.south east) -- (n22.north west)
    ;  
\draw[dashed, gray] 
     (n13.north east) -- (n23.south west)
    (n13.south east) -- (n23.north west)
    ;  
    
\draw[dashed, gray] 
    ([shift={(1ex,-1ex)}] r2.south east) -- ( n11.south west)
    ([shift={(1ex,+4ex)}] r2.south) -- ( n11.north west)
    
    ([shift={(1ex,+1ex)}] r2.south west) -- ( n13.north west)
    ([shift={(1ex,-4ex)}] r2.south) -- ( n13.south west)
    
    ([shift={(1ex,-3ex)}] r2.south) -- ( n12.north west)
    ([shift={(1ex,+3ex)}] r2.south) -- (n12.south west);
\end{tikzpicture}
\caption{Multi task learning framework}
\label{fig.mtl}
\end{figure}
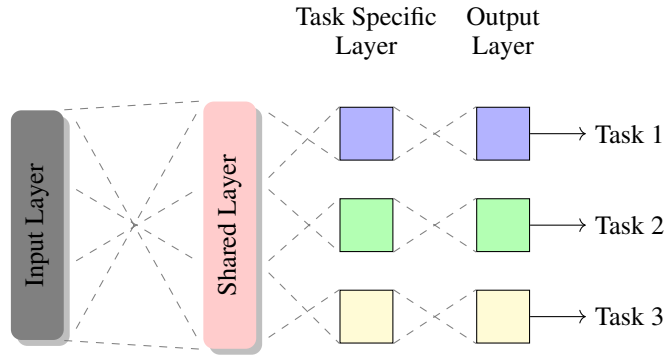
Figure \ref{fig.mtl} shows MTL framework used in our problem. The shared layers contain the following layers:  a fully connected layer with $20$ hidden units, dropout layer with rate $0.2$, ReLU, a fully-connected layer with $10$ hidden units. Below are the layer detail for each task-
\begin{itemize}[leftmargin=5.5mm]
    \item Age prediction (Task 1)- A fully connected layer with $10$ hidden units, ReLU, one output unit
    \item Site prediction (Task 2)- A fully connected layer with $10$ hidden units, ReLU, a softmax layer
    \item Sex prediction (Task 3)-  A fully connected layer with $10$ hidden units, ReLU, a softmax layer
\end{itemize}

\end{document}